\documentclass[a4paper,twocolumn,english,final,lengthcheck,showpacs,pra,superscriptaddress]{revtex4}
\usepackage[T1]{fontenc}
\usepackage[latin9]{inputenc}
\usepackage{amsmath}
\usepackage{graphicx}
\usepackage{amssymb}

\makeatletter

\providecommand{\tabularnewline}{\\}

\@ifundefined{textcolor}{}
{%
 \definecolor{BLACK}{gray}{0}
 \definecolor{WHITE}{gray}{1}
 \definecolor{RED}{rgb}{1,0,0}
 \definecolor{GREEN}{rgb}{0,1,0}
 \definecolor{BLUE}{rgb}{0,0,1}
 \definecolor{CYAN}{cmyk}{1,0,0,0}
 \definecolor{MAGENTA}{cmyk}{0,1,0,0}
 \definecolor{YELLOW}{cmyk}{0,0,1,0}
 }

\usepackage{amsfonts}

\providecommand{\U}[1]{\protect\rule{.1in}{.1in}}
\providecommand{\U}[1]{\protect\rule{.1in}{.1in}}
\providecommand{\U}[1]{\protect\rule{.1in}{.1in}}

\makeatother

\usepackage{babel}

\begin{document}

\title{Decoherence as attenuation of mesoscopic echoes in a spin-chain channel}

\author{Gonzalo A. \'{A}lvarez}

\email{galvarez@e3.physik.uni-dortmund.de}

\affiliation{Fakult\"{a}t Physik, Technische Universit\"{a}t Dortmund, D-44221 Dortmund,
Germany.}

\author{Ernesto P. Danieli}

\email{edanieli@mc.rwth-aachen.de}

\affiliation{ITMC, RWTH Aachen University, D-52074 Aachen, Germany.}

\author{Patricia R. Levstein}

\affiliation{Facultad de Matem\'{a}tica, Astronom\'{\i}a y F\'{\i}sica and Instituto de F\'{\i}sica
Enrique Gaviola, Universidad Nacional de C\'{o}rdoba, 5000 C\'{o}rdoba, Argentina.}

\author{Horacio M. Pastawski}

\email{horacio@famaf.unc.edu.ar}

\affiliation{Facultad de Matem\'{a}tica, Astronom\'{\i}a y F\'{\i}sica and Instituto de F\'{\i}sica
Enrique Gaviola, Universidad Nacional de C\'{o}rdoba, 5000 C\'{o}rdoba, Argentina.}


\begin{abstract}
An initial local excitation in a confined quantum system evolves exploring
the whole system, returning to the initial position as a mesoscopic
echo at the Heisenberg time. We consider a two weakly coupled spin
chains, a spin ladder, where one is a quantum channel while the other
represents an environment. We quantify decoherence in the quantum
channel through the attenuation of the mesoscopic echoes. We evaluate
decoherence rates for different ratios between sources of amplitude
fluctuation and dephasing in the inter-chain interaction Hamiltonian.
The many-body dynamics is seen as a one-body evolution with a decoherence
rate given by the Fermi golden rule.
\end{abstract}
\maketitle

\section{Introduction}

Control of quantum dynamics is essential to achieve quantum information
processing. Examples of this are implementations of quantum algorithms
\cite{z--Grover,z-Shor} and quantum communications \cite{z--teleportI,z--teleportII}.
The system involved in each of these processes interacts with an environment
which degrades the quantum correlations \cite{z--Zurek}. This loss
of information, called decoherence, represents the main obstacle to
achieve an efficient quantum processing. Thus, in order to avoid decoherence,
it is mandatory to understand its processes. Many approaches exist
to study decoherence in systems composed by few qubits \cite{Koppens06,Wu06,Grajcar06,Riebe06}.
However, real computer implementations that involve large number of
qubits require the development of new approaches to characterize decoherence
\cite{Krojanski04,Krojanski06,AlvarezPRL10}.

Recently, many efforts were done to characterize the quantum noise
\cite{Chuang00,z--Guo,z--BoseDeco} 
that produces decoherence in quantum channels. These channels connect
two quantum systems enabling the information transfer between them.
Spin chains can be used to achieve this goal for short distance communications
\cite{z--Bose} avoiding interfaces between the static system and
the information carriers. Whereas all these studies were oriented
to pure-state communication processes, experimental realizations of
pure-state dynamics are a major challenge \cite{z--QCRoadmap04}.
Alternatively, implementations of quantum computation with NMR can
not deal with pure states but have to resort to statistical mixtures
of spin ensembles instead \cite{Cory97,Chuang97,Laflame98}. State
transfer in spin ensembles was done in a ring of spins with many-body
interactions in the solid state \cite{z--EcosMesoscI,z--EcosMesoscII}.
An initial local polarization propagates around the ring. The return
of the initial excitation is evidenced through the constructive interference
that reappears at the Heisenberg time $t_{\mathrm{H}}\sim\hbar/\Delta$,
with $\Delta$ the typical mean energy level spacing, as a form of
polarization revival called the mesoscopic echo \cite{z--ME-Altshuler}.
There, the polarization amplitude and phase of a given nuclear spin
within the ring was monitored as a function of time. However, the
many body nature of spin-spin interactions strongly compromises an
optimum transfer. Thus, a much more efficient transfer was observed
through the experimental implementation of an effective $XY$ Hamiltonian
(i.e. flip-flop processes) in a spin chain \cite{z--Madi} by exploiting
the $J$-coupling in the liquid phase, where dipolar interaction becomes
negligible. The main reason for this result is that the many-body
dynamics of an $XY$ Hamiltonian is mappable to a one-body evolution
\cite{Lieb61}. Thus, in this case, the Heisenberg time is proportional
to the system size $M$ instead of the $2^{M}$ value that shows in
a complex many-body Hamiltonian. 
In Ref. \cite{z--Madi}, the evolution of the initial excitation was
monitored in all the spins of the quantum channel. Comparisons with
theoretical calculations showed the effect of decoherence manifested
in the attenuation of the interference intensities which is more evident
in the decay of the mesoscopic echoes. This experimental breakthrough
enabled various theoretical proposals for perfect state transfer \cite{Christandl04,Stolze05}.
More recently, implementations of spin-chains were done in solid-state
NMR \cite{CapellaroPRA07} by implementing a Double Quantum Hamiltonian
(i.e. flip-flip+flop-flop processes) which, in turn, is mappable to
an XY interaction \cite{Doronin}. However, the progress toward the
proposal to observe the mesoscopic echoes of these systems \cite{CapellaroPRL07}
was experimentally limited by length inhomogeneity \cite{Rufeil-Fiori-MQC}.

In this work we propose to use the attenuation of the mesoscopic echoes
in a spin chain as a sensor of the decoherence produced by an uncontrolled
spin bath, in this case a second spin chain. The spin-spin interaction
within each chain is given by an $XY$ Hamiltonian where the ensemble
dynamics can be solved analytically \cite{z--FeldmanErnst,z-SPC,z--NuestroCPL05}.
Once the chains are laterally coupled to form a spin ladder, the quantum
dynamics becomes truly many-body and the analytical solution is no
longer possible. Moreover, numerical solutions are difficult to obtain
as a consequence of the exponential increase of the Hilbert space
dimension \cite{Quantum-Paralelism}. We show that, within certain
range of the ratio between the inter-chain and intra-chain interactions,
the evolution of a local excitation in the many-body system (spin
ladder) can be obtained as a one-body dynamics (isolated chain evolution)
plus a decoherence process given by a Fermi golden rule (FGR). We
characterize the decoherence rate for different kinds of inter-chain
interactions by controlling the contribution of the different sources:
pure dephasing (Ising interaction) and amplitude fluctuations ($XY$
interaction). Thus, we show that this auto-correlation function becomes
an effective and practical tool for studying decoherence. Moreover,
the exploration nature of the mesoscopic echo provides the sender
(Alice) with the information of the decoherence effects in the quantum
channel by contrasting with the expected result of the isolated dynamics.
This could enable the implementation of a convenient error correction
code without needing of a receiver.

In the next section, we introduce the system and present the numerical
calculations of the local polarization evolution for different inter-chain
interactions. From these results we obtain the decoherence rates from
the attenuation of the mesoscopic echoes. In the third section, for
a solvable small system, we characterize analytically the decoherence
rates. Instead of using the quantum master equation which is the most
standard framework adopted to describe the system-environment interaction
\cite{z--QmasterE1,z--QmasterE2}, we use the Keldysh non-equilibrium
formalism which leads to an integral solution of the Schrödinger equation
\cite{z-SPC,z--NuestroCPL05,z--NuestroSSC07,z--NuestroPRA07}. Finally,
we discuss the conclusions.

\section{Spin dynamics in a non-isolated spin-chain channel\label{sec I}}

\subsection{System\label{sec 1.1}}

\begin{figure}[tbh]

\begin{centering}
\includegraphics[width=3.3062in,height=2.2035in]{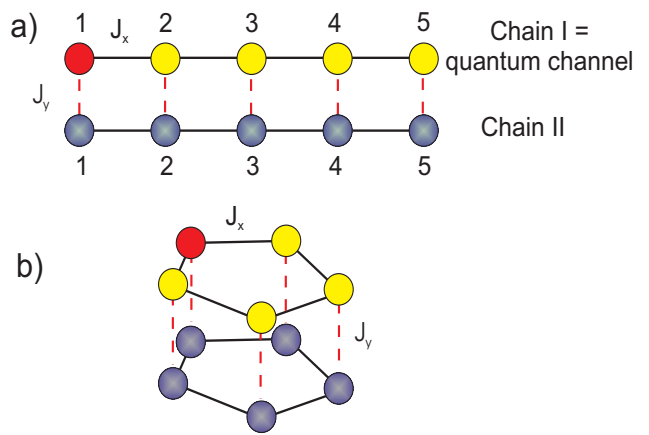} 
\par\end{centering}

\caption{(Color online): (a) Schematic representation of the spin ladder system
given by the Hamiltonian of Eq. (\ref{eq--Htotal}). Each spin chain
has $M=5$ spins. (b) A system like in (a) but with periodic boundary
conditions between spins $1$ and $M$. Light colors represent the
quantum channel. Marked spin (red online) represents the site where
the initial condition is placed. }

\label{f--figure1}
\end{figure}

Let us consider the polarization dynamics in a spin system composed
by two parallel spin chains transversely coupled. The topology of
the interaction network forms a ladder. Chain I represents the quantum
channel composed by $M$ spins, and chain II is another $M$ spin
system that perturbs the quantum transfer in chain I, see Fig. \ref{f--figure1}(a).
The spin Hamiltonian is given by\begin{equation}
\widehat{\mathcal{H}}_{\mathrm{total}}=\widehat{\mathcal{H}}_{\mathrm{I}}+\widehat{\mathcal{H}}_{\mathrm{II}}+\widehat{\mathcal{H}}_{\mathrm{T}},\label{eq--Htotal}\end{equation}
 where \begin{align}
\widehat{\mathcal{H}}_{\mathrm{\alpha}} & =\sum_{n=1}^{M-1}J_{n+1,n}^{(\mathrm{\alpha})}\left(\widehat{S}_{n+1}^{_{(\mathrm{\alpha})}x\ }\widehat{S}_{n}^{_{(\mathrm{\alpha})}x}+\widehat{S}_{n+1}^{_{(\mathrm{\alpha})}y\ }\widehat{S}_{n}^{_{(\mathrm{\alpha})}y}\right)\label{eq--Chain_Hamiltonian}\\
 & =\sum_{n=1}^{M-1}\frac{J_{n+1,n}^{(\mathrm{\alpha})}}{2}\left(\widehat{S}_{n+1}^{_{(\mathrm{\alpha})}+\ }\widehat{S}_{n}^{_{(\mathrm{\alpha})}-}+\widehat{S}_{n+1}^{_{(\mathrm{\alpha})}+\ }\widehat{S}_{n}^{_{(\mathrm{\alpha})}-}\right)\end{align}
 represents the $\mathrm{\alpha}$-th spin chain Hamiltonian that
takes into account the $XY$ interaction between neighbor 1/2-spins
within the chain. The spin chains interact through the following transversal
Hamiltonian\begin{align}
\widehat{\mathcal{H}}_{\mathrm{T}} & =\sum_{n=1}^{M}aJ_{n}^{{}}\widehat{S}_{n}^{_{(\mathrm{I})}z}\widehat{S}_{n}^{_{(\mathrm{II})}z}+bJ_{n}^{{}}\left(\widehat{S}_{n}^{_{(\mathrm{I})}x}\widehat{S}_{n}^{_{(\mathrm{II})}x}+\widehat{S}_{n}^{_{(\mathrm{I})}y}\widehat{S}_{n}^{_{(\mathrm{II})}y}\right)\label{eq--Tranverse H}\\
 & =\sum_{n=1}^{M}aJ_{n}^{{}}\widehat{S}_{n}^{_{(\mathrm{I})}z}\widehat{S}_{n}^{_{(\mathrm{II})}z}+\frac{bJ_{n}}{2}\left(\widehat{S}_{n}^{_{(\mathrm{I})}+}\widehat{S}_{n}^{_{(\mathrm{II})}-}+\widehat{S}_{n}^{_{(\mathrm{I})}+}\widehat{S}_{n}^{_{(\mathrm{II})}-}\right),\end{align}
 where by controlling the ratio $a/b$ one determines the nature of
the interaction. The polarization evolution is considered for different
transversal interactions: $XY$ interaction ($a/b=0$); isotropic
(Heisenberg) interaction ($a/b=1$); and truncated dipolar interaction
($a/b=-2$), which are typical in NMR experiments \cite{z--QmasterE1,z--QmasterE2}.
Besides these three typical NMR interactions we will also consider
the evolution under two additional transversal Hamiltonian with the
following parameters $H_{1}$ ($a/b=2$) and $H_{2}$ ($a/b=3.6$).
We will focus on regular systems where all longitudinal couplings
within chains are taken equals $J_{n+1,n}^{(\mathrm{\alpha})}=J_{\mathrm{x}}$
$(n\neq M)$. The same is applied to the transversal couplings $J_{n}=J_{\mathrm{y}}$.
We choose these parameter conditions because the interference effects
are more pronounced and the mesoscopic echo degradation can be easily
evaluated. 
%

\subsection{Decoherence characterization based on the attenuation of mesoscopic
echoes}

In order to study the mesoscopic echoes of a spin ensemble, we calculate
the evolution of a local polarization within chain I (quantum channel)
through the spin auto-correlation function \cite{z-SPC,z--NuestroCPL05},
\begin{equation}
P_{1,1}(t)=\frac{\left\langle \Psi_{\mathrm{eq}}\right\vert \widehat{S}_{1}^{_{(\mathrm{I})}z}(t)\widehat{S}_{1}^{_{(\mathrm{I})}z}(0)\left\vert \Psi_{\mathrm{eq}}\right\rangle }{\left\langle \Psi_{\mathrm{eq}}\right\vert \widehat{S}_{1}^{_{(\mathrm{I})}z}(0)\widehat{S}_{1}^{_{(\mathrm{I})}z}(0)\left\vert \Psi_{\mathrm{eq}}\right\rangle }.\label{eq--PCFunc}\end{equation}
 This gives the local polarization in the $z$ direction on site $1$
at time $t$ providing that the system was in its equilibrium state
plus a local excitation on site $1$ at time $t=0$. Here, $\widehat{S}_{1}^{_{(\mathrm{\alpha})}z}\left(t\right)=e^{\mathrm{i}\widehat{\mathcal{H}}_{\mathrm{total}}t/\hbar}\widehat{S}_{1}^{_{(\mathrm{\alpha})}z}e^{-\mathrm{i}\widehat{\mathcal{H}}_{\mathrm{total}}t/\hbar}$
is the spin operator in the Heisenberg representation and $\left\vert \Psi_{\mathrm{eq}}\right\rangle $
is the many-body state corresponding to thermal equilibrium, that
is a mixture with appropriate statistical weights, of all possible
states with different number $N$ of spins up. In the regime of NMR
spin dynamics $k_{\mathrm{B}}T$ is much\ higher than any energy
scale of the system. Then, all the statistical weights can be taken
equal \cite{z--QmasterE1,z--QmasterE2}. As numerical method we alternate
between\ the standard diagonalization by spin projection subspaces
\cite{z--EcosMesoscI} or the Quantum Parallelism algorithm \cite{Quantum-Paralelism},
which involves the evolution of a few superposition states representing
the whole ensemble. Although the last is much more efficient for larger
samples, for the considered system sizes, both involve similar computing
time. As the results coincide, no further comment is devoted to this
point.

\begin{figure}[tbh]
\begin{centering}
\includegraphics[width=2.8781in,height=4.1883in]{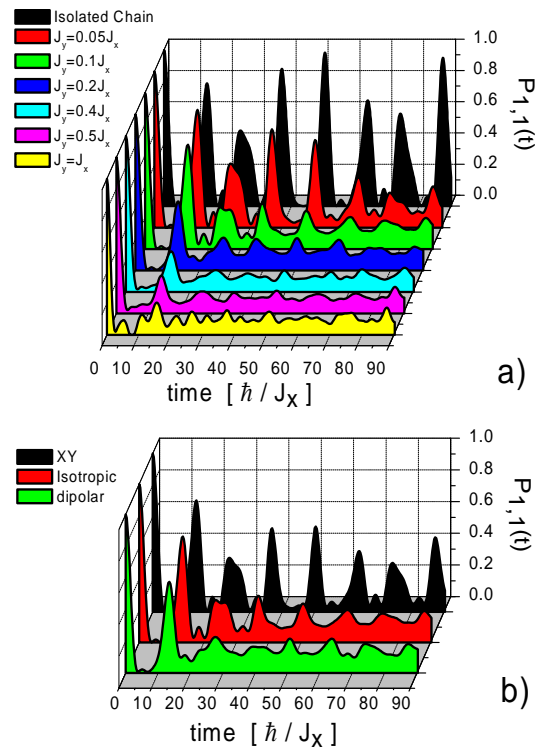}
\par\end{centering}

\caption{(Color online): Polarization at site $1$, $P_{1,1}(t)$, of the spin
ladder of Fig \ref{f--figure1}(a). (a) Upper (black) curve represents
the dynamics of an isolated chain given by $P_{1,1}^{\text{\textrm{isolated}}}(t)$.
Color curves correspond to different values of the coupling $J_{\mathrm{y}}$
for a transversal isotropic interaction between chains. (b) Comparison
of the dynamics behavior between different transversal interaction
natures ($XY$, Isotropic, dipolar) by keeping fixed the coupling
ratio $J_{\mathrm{y}}/J_{\mathrm{x}}=0.1$. }

\centering{}\label{f--figure2}
\end{figure}

By assuming $J_{\mathrm{y}}$ equal to zero, the local polarization
dynamics at spin $1$ in chain I is equivalent to the local evolution
in an isolated spin chain, $P_{1,1}^{\text{\textrm{isolated}}}(t)$.
This function is defined with Eq. (\ref{eq--PCFunc}) replacing $\widehat{\mathcal{H}}_{\mathrm{total}}$
by $\widehat{\mathcal{H}}_{\mathrm{I}}$ and $\widehat{S}_{1}^{_{(\mathrm{I})}z}\left(t\right)=e^{\mathrm{i}\widehat{\mathcal{H}}_{\text{\textrm{I}}}t/\hbar}\widehat{S}_{1}^{_{(\mathrm{I})}z}e^{-\mathrm{i}\widehat{\mathcal{H}}_{\text{\textrm{I}}}t/\hbar}$.
Since the $XY$ spin-spin interaction within the chain only couples
nearest neighbors, the system dynamics in the high temperature regime
can be obtained analytically as a one-body dynamics \cite{z-SPC,z--NuestroCPL05}.
Upper line (black) in Figure \ref{f--figure2}(a) shows the solution
of $P_{1,1}^{\text{\textrm{isolated}}}(t)$ for a spin chain with
$M=5$. We can observe the presence of the $k$-th mesoscopic echo
at a time $t_{k}^{\text{\textrm{ME}}}$ proportional to the chain
size $M$ \cite{z--EcosMesoscI}. As $J_{\mathrm{y}}$ increases,
the degrees of freedom of chain II start modifying the observed dynamics
of $P_{1,1}^{\text{\textrm{isolated}}}(t)$ and, in general, no analytical
solution is known. To obtain the polarization dynamics $P_{1,1}(t)$
of this many-body problem, we solve numerically the time dependent
Schrödinger equation by using the Trotter-Suzuki decomposition \cite{z--DeRaedt}.
Color curves of Fig. \ref{f--figure2}(a) show the local polarization
$P_{1,1}(t)$ for different values of $J_{\mathrm{y}}$ corresponding
to an isotropic transversal Hamiltonian. They evidence the degradation
of mesoscopic echoes proportionally to $J_{\mathrm{y}}$. Fig. \ref{f--figure2}(b)
shows $P_{1,1}(t)$ for different forms of the transversal interaction
Hamiltonian for a fixed value of $J_{\mathrm{y}}$. Due to the $XY$
term in the transversal Hamiltonian, the polarization is transferred
back and forth between chain I and II. This process, plus the dephasing
induced by the Ising contribution, produce the progressive spreading
of the polarization among all the spins thus degrading the strong
recurrences. This is manifested in the mean value of the local polarization
at longer times, $t\gg\hbar/J_{\mathrm{x}},$ that approximately tends
to $1/\left(2M\right)$. Thus, one observes a decrease of the mesoscopic
echoes with evolution time as well as a gradual increase of the background
polarization at times between echoes. We do not consider an Ising
transversal interaction ($b/a=0$) since it does not involve transfer
of polarization between chains leading to a different value of the
mean local magnetization at long times. In the latter case, the mean
local polarization at long times tends to a value that is close to
$1/M$ instead of the $1/\left(2M\right)$ value. This difference
in the final state avoid a direct comparison with the polarization
dynamics from Hamiltonians that contain an $XY$ term requiring extra
manipulations.
%


In order to characterize decoherence of the quantum channel, we measure
the attenuation of the mesoscopic echoes as compared with the local
polarization in the isolated spin-chain channel, $P_{1,1}^{\text{\textrm{isolated}}}(t)$.
Figure \ref{f--figure2} shows that the ratio $P_{1,1}^{{}}(t_{k}^{\text{\textrm{ME}}})/P_{1,1}^{\text{\textrm{isolated}}}(t_{k}^{\text{\textrm{ME}}})$
decreases as a function of $t_{k}^{\text{\textrm{ME}}}$. Within the
regime $\left\vert J_{\mathrm{y}}/J_{\mathrm{x}}\right\vert \ll1$
we observe that this ratio has an exponential dependence of the form
$\exp[-t_{k}^{\text{\textrm{ME}}}/\tau_{\phi}]$. This exponential
decay of the initial polarization cannot hold for all times until
$P_{1,1}^{{}}(t)=0$ since the system size is finite and the total
polarization is conserved within the system. Thus, for a later time
the correlations within chain II are manifested at site $1$, and
the ratio $P_{1,1}^{{}}(t_{k}^{\text{\textrm{ME}}})/P_{1,1}^{\text{\textrm{isolated}}}(t_{k}^{\text{\textrm{ME}}})$
evidences a complex behavior instead of an exponential one.\ The
numerical results show that the crossover between the two temporal
behavior appears at times proportional to $t_{\mathrm{R}}\propto\tau_{\phi}\ln[J_{x}\tau_{\phi}/\hbar]$
\cite{z--elenaCPL05}. To characterize the exponential decay of the
mesoscopic echoes we calculate the ratio $P_{1,1}^{{}}(t_{k}^{\text{\textrm{ME}}})/P_{1,1}^{\text{\textrm{isolated}}}(t_{k}^{\text{\textrm{ME}}})$
for several values of $t_{k}^{\text{\textrm{ME}}}$ previous to this
crossover time. The accuracy of this characterization improves proportionally
to the order $k$ of the mesoscopic echo. Thus, to avoid dealing with
too long times or large number of spins, we increase the order of
observable echoes without changing the physical behavior of the system
by putting periodic boundary condition in $\widehat{\mathcal{H}}_{\mathrm{I,II}}$.
This {}``closes\textquotedblright{}\ the spin chain into a ring
as shown in figure \ref{f--figure1}(b). This choice approximately
duplicate the number of mesoscopic echoes for a fixed evolution time
with respect to the observed in the chain.

\begin{figure}[tbh]
\begin{centering}
\includegraphics[width=3.3018in,height=2.6377in]{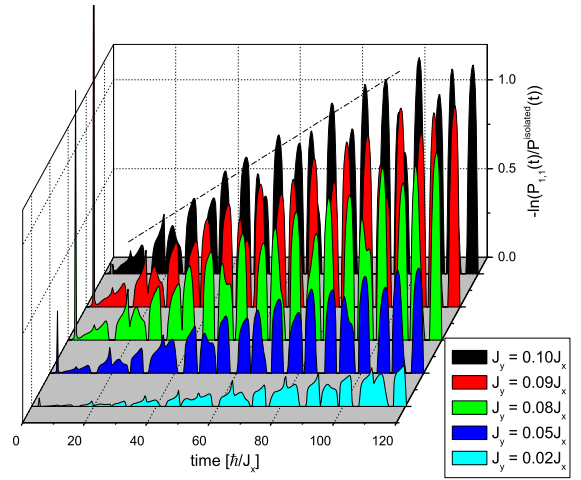}
\par\end{centering}

\caption{(Color online): Natural logarithm of the ratio $P_{11}(t)/P_{1,1}^{\text{\textrm{isolated}}}(t)$
for different values of the transversal coupling $J_{\mathrm{y}}$
as a function of time. The interaction between chains has an isotropic
nature. The dashed line is a guide to the eye.}

\centering{}\label{f--figure3}
\end{figure}

\begin{figure}[tbh]
\begin{centering}
\includegraphics{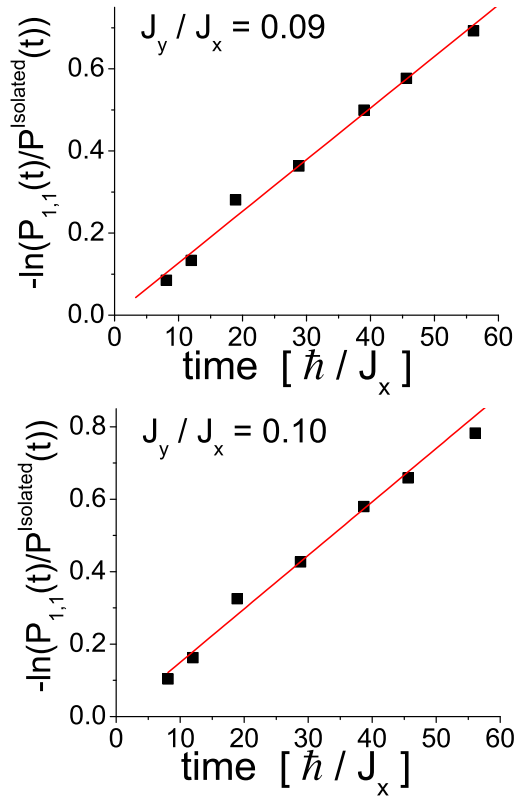}
\par\end{centering}

\caption{(Color online): Maximum values of Fig. \ref{f--figure3} for two different
ratios $J_{\mathrm{y}}/J_{\mathrm{x}}$ as a function of time. The
slope of the linear fit (solid line) gives the value of $\tau_{\phi}$
associated to the perturbation $J_{\mathrm{y}}$.}

\centering{}\label{f--figure3-1}
\end{figure}

Since chain II does not have infinite degrees of freedom it does not
represent a reservoir. Thus, the presence of recurrences in the quotient
$P_{1,1}^{{}}(t_{k}^{\text{\textrm{ME}}})$/$P_{1,1}^{\mathrm{isolated}}(t_{k}^{\text{\textrm{ME}}})$
is expected instead of an exponential law. In figure \ref{f--figure3},
the function $-\log(P_{1,1}^{{}}(t)/P_{1,1}^{\mathrm{isolated}}(t))$
is plotted for different values of $J_{\mathrm{y}}$ as a function
of time. Due to the dynamics within the channel each curve oscillates
around zero. The maximum values are associated to the temporal region
where $P_{1,1}^{\mathrm{isolated}}(t)>P_{1,1}(t)$, that is, the region
where the mesoscopic echoes are manifested. The minimum values, on
the other hand, correspond to the temporal regions between mesoscopic
echoes, where $P_{1,1}^{\mathrm{isolated}}(t)<P_{1,1}(t)$. It can
be seen that an exponential law is identified in the envelop of the
peaks that appears for each curve (see dashed line). In figure \ref{f--figure3-1}
are plotted the values of the maxima of figure \ref{f--figure3} for
$J_{\mathrm{y}}=0.09J_{\mathrm{x}}$ and $J_{\mathrm{y}}=0.10J_{\mathrm{x}}$
as a function of time. The peaks start to get apart from its envelop
at different times proportional to $t_{\mathrm{R}}$ depending on
the intensities of the transverse interaction. This behavior reflects
the recurrences due to the finite nature of chain II, which becomes
relevant for earlier times as the intensities $J_{\mathrm{y}}$ increases.
In order to quantify the exponential behavior corresponding to each
value of the transverse interaction $J_{\mathrm{y}}$, we fit the
maximum values shown in Fig. \ref{f--figure3-1} to a linear function.
Thus, we extract the decoherence time $\tau_{\phi}$ from the slope
of the curve for different values of the transversal interaction $J_{\mathrm{y}}$.
The number of points (number of echoes) used in this procedure varies
depending on the value of $J_{\mathrm{y}}$ as can be seen in Fig.
\ref{f--figure3-1}. Besides, the point $(0,0)$ is not included in
the fit since for very short times $t\ll\hbar/J_{\mathrm{x}}$ the
behavior of quantum dynamics is not exponential but it starts quadratic
with $t$.
%

%

The existence of an exponential law in the decay of the mesoscopic
echoes where $\tau_{\phi}\propto\left\vert \hbar J_{\mathrm{x}}/J_{\mathrm{y}}^{\ 2}\right\vert $
plus the fact that its time regime is bounded by a time proportional
to $t_{\mathrm{R}}\propto\tau_{\phi}\ln[J_{x}\tau_{\phi}/\hbar]$
suggests that the decay of mesoscopic echoes in chain I could be described
by the Self Consistent Fermi Golden Rule (SC-FGR) analyzed in Ref.
\cite{z--elenaCPL05}.

\begin{figure}[tbh]
\begin{centering}
\includegraphics[clip,width=3.3823in,height=4.3561in]{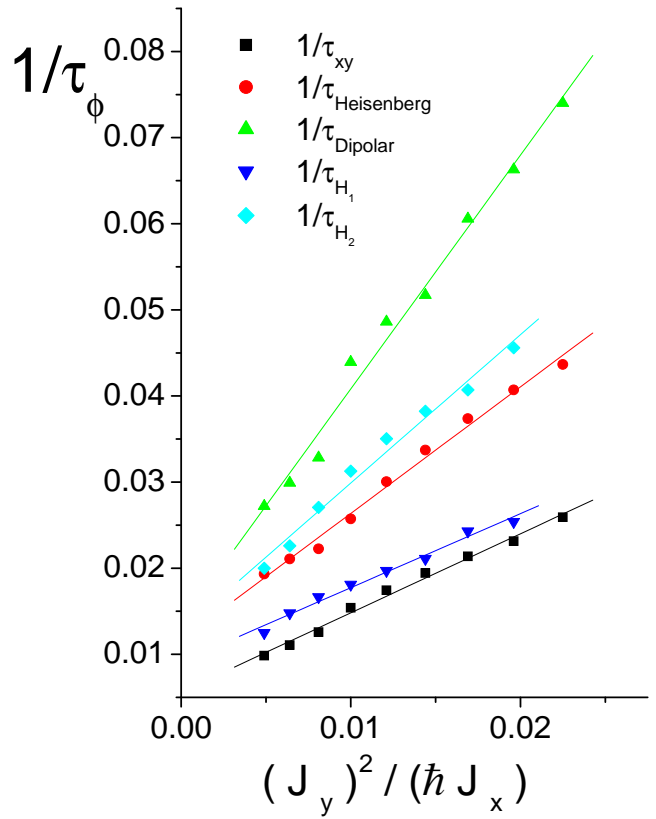}
\par\end{centering}

\caption{(Color online): Decoherence rates $\frac{1}{\tau_{\phi}}$ for different
transversal Hamiltonians as a function of $\left\vert J_{\mathrm{y}}\right\vert ^{2}/\hbar J_{\mathrm{x}}$.
The lineal dependence indicates the good agreement with the FGR behavior.
Different slopes indicate that the decoherence rate is composed by
two contributions: One is due to the injection of polarization produced
by the $XY$ term of $\widehat{\mathcal{H}}_{\mathrm{T}}$ and the
other is associated to the dephasing process given by the Ising term.}

\centering{}\label{f--figure4}
\end{figure}

Figure \ref{f--figure4} shows the obtained values of $\tau_{\phi}$
as a function of $\frac{\left\vert J_{\mathrm{y}}\right\vert ^{2}}{\hbar J_{\mathrm{x}}}$
evidencing a linear dependence. The curves correspond to transversal
interaction of different nature by varying the ratio $a/b$ in Eq.
(\ref{eq--Tranverse H}): $XY$ ($a/b=0$), isotropic or Heisenberg
($a/b=1$), truncated dipolar ($a/b=-2$), $H_{1}$ ($a/b=2$) and
$H_{2}$ ($a/b=3.6$). The slopes of the linear curves are shown in
table \ref{TableKey}.
%

%
\begin{table}
\centering 
\begin{tabular}{|c|c|}
\hline 
$\widehat{\mathcal{H}}_{\mathrm{T}}$  & $\frac{1}{\tau_{\phi}}$\tabularnewline
\hline 
$XY$ ($a/b=0$)  & $\left(0.96\pm0.04\right)\frac{\left\vert J_{\mathrm{y}}\right\vert ^{2}}{\hbar J_{\mathrm{x}}}$\tabularnewline
\hline 
Isotropic ($a/b=1$)  & $\left(1.47\pm0.05\right)\frac{\left\vert J_{\mathrm{y}}\right\vert ^{2}}{\hbar J_{\mathrm{x}}}$\tabularnewline
\hline 
dipolar ($a/b=-2$)  & $\left(2.7\pm0.1\right)\frac{\left\vert J_{\mathrm{y}}\right\vert ^{2}}{\hbar J_{\mathrm{x}}}$\tabularnewline
\hline 
$H_{1}$ ($a/b=2$)  & $\left(0.81\pm0.04\right)\frac{\left\vert J_{\mathrm{y}}\right\vert ^{2}}{\hbar J_{\mathrm{x}}}$\tabularnewline
\hline 
$H_{2}$ ($a/b=3.6$)  & $\left(1.8\pm0.1\right)\frac{\left\vert J_{\mathrm{y}}\right\vert ^{2}}{\hbar J_{\mathrm{x}}}$\tabularnewline
\hline
\end{tabular}

\caption{Proportional constants between $1/t_{\phi}$ and $|J_{y}|^{2}/(\hbar J_{x})$
for different interactions between chains.}

\label{TableKey}

\end{table}


The linear behavior observed in figure \ref{f--figure4} confirms
that, even when we are dealing with closed systems in which the energetic
spectrum of the environment (chain II) is not continuous, the characteristic
time $\tau_{\phi}$ agrees with the one dictated by the FGR. The conditions
needed to derive the FGR in molecular spin systems was already addressed
in ref. \cite{z--elenaCPL05}. There, the behavior of the numerical
simulations on clusters with less than $20$ spins for times shorter
than $t_{\mathrm{H}}$ was observed to coincide perfectly with the
theoretical results.

\begin{figure}[tbh]

\begin{centering}
\includegraphics[width=3.0874in,height=4.3128in]{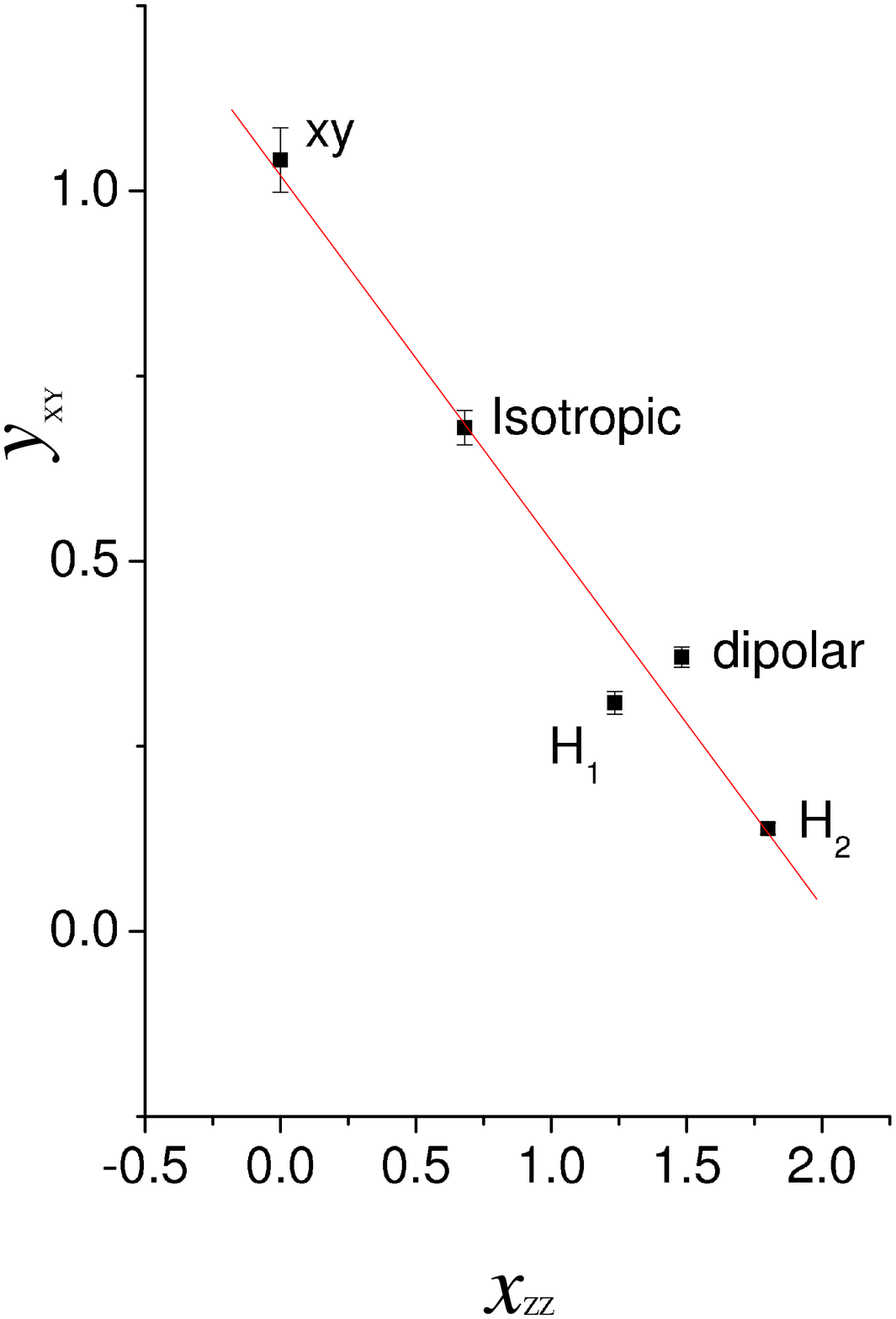}
\par\end{centering}

\caption{(Color online): $y_{\mathrm{XY}}^{{}}$ vs. $x_{\mathrm{ZZ}}^{{}}$
for different natures of the transversal interaction, where $x_{\mathrm{ZZ}}^{{}}=a^{2}\tau_{\phi}\left\vert J_{\mathrm{y}}\right\vert ^{2}/\left(\hbar J_{\mathrm{x}}\right)$
and $y_{\mathrm{XY}}^{{}}=b^{2}\tau_{\phi}\left\vert J_{\mathrm{y}}\right\vert ^{2}/\hbar J_{\mathrm{x}}$.
The solid line is given by a linear fitting to the points. }

\centering{}\label{f--figure5}
\end{figure}

In the present case, $5+5$ spins, the unperturbed Hamiltonian is
represented by $\widehat{\mathcal{H}}_{\mathrm{I}}+\widehat{\mathcal{H}}_{\mathrm{II}}$,
while the perturbation is $\widehat{\mathcal{H}}_{\mathrm{T}}$. The
initial state is not an eigenstate of $\widehat{\mathcal{H}}_{\mathrm{I}}+\widehat{\mathcal{H}}_{\mathrm{II}}$,
but instead it is a superposition of the $2^{5}$ eigenstates of chain
I. Since the initial state is spatially localized, all the eigenstates
of chain I contribute with equal weight to this superposition. Therefore,
the decay of a local polarization involves a sort of average over
all possible initial and final states. This justifies the use of the
expression for the characteristic time given by the Fermi Golden Rule
\cite{Pascazio99} \begin{equation}
\frac{1}{\tau_{\phi}}\simeq\frac{2\pi}{\hbar}\left\Vert \widehat{\mathcal{H}}_{\mathrm{T}}\right\Vert ^{2}N_{0},\label{eq--fgr}\end{equation}
 where $\left\Vert \widehat{\mathcal{H}}_{\mathrm{T}}\right\Vert $
is a characteristic value of the coupling interactions between chains
that has to be determined and $N_{0\text{ }}$represents the density
of directly connected states, i.e. states of the states of chain II
connected with the initial state through the perturbation $\widehat{\mathcal{H}}_{\mathrm{T}}$.
This interaction consists of two processes: the flip-flop (or $XY)$
interaction and the Ising interaction weighted by the factors $a$
and $b$ respectively. Thus, the evolution of the initial localized
state generated in chain I decays due to the coupling with the {}``environment\textquotedblright{}\ represented
by chain II. Equation (\ref{eq--fgr}) involves the assumption that
$N_{0\text{ }}$is similar for the $XY$ and Ising interaction, i.e.,
$N_{0}^{\mathrm{XY}}=N_{0}^{\mathrm{ZZ}}=N_{0}$. This density of
states can be approximated as the inverse of the second moment of
$\widehat{\mathcal{H}}_{\text{\textrm{II}}}$, and then $N_{0}\propto1/J_{\mathrm{x}}$.
The cross terms proportional to $ab$ are canceled out in $\left\Vert \widehat{\mathcal{H}}_{\mathrm{T}}\right\Vert ^{2}$
\cite{z--QmasterE1}, thus, under these approximations each interaction
terms between chains becomes independent of each other obtaining\begin{equation}
\frac{1}{\tau_{\phi}^{{}}}=\frac{1}{\tau_{\phi}^{\mathrm{ZZ}}}+\frac{1}{\tau_{\phi}^{\mathrm{XY}}},\label{eq--taum1taum1XYtaum1I}\end{equation}
 with \begin{equation}
\frac{1}{\tau_{\phi}^{\mathrm{ZZ}}}=Aa^{2}\frac{\left\vert J_{\mathrm{y}}\right\vert ^{2}}{\hbar J_{\mathrm{x}}}\text{ and\textrm{ }}\frac{1}{\tau_{\phi}^{\mathrm{XY}}}=Bb^{2}\frac{\left\vert J_{\mathrm{y}}\right\vert ^{2}}{\hbar J_{\mathrm{x}}},\label{eq--taum1Iandtaum1XY}\end{equation}
 where $A$ and $B$ represent constants associated with each interaction.
We determine these constants by using the numerical results of table
\ref{TableKey}. Equation (\ref{eq--taum1taum1XYtaum1I}) can be rewritten
as\begin{equation}
y_{\mathrm{XY}}^{{}}=-\frac{A}{B}x_{\mathrm{ZZ}}^{{}}+\frac{1}{B}\end{equation}
 where $x_{\mathrm{ZZ}}^{{}}=a^{2}\tau_{\phi}\left\vert J_{\mathrm{y}}\right\vert ^{2}/\left(\hbar J_{\mathrm{x}}\right)$
and $y_{\mathrm{XY}}^{{}}=b^{2}\tau_{\phi}\left\vert J_{\mathrm{y}}\right\vert ^{2}/\left(\hbar J_{\mathrm{x}}\right)$.
Figure (\ref{f--figure5}) shows the curve $y_{\mathrm{XY}}^{{}}$
vs. $x_{\mathrm{ZZ}}^{{}}$ (squares) for the three standard NMR Hamiltonians,
$XY$ ($b=1$, $a=0$), Isotropic ($b=1$, $a=1$), dipolar ($b=1,a=-2$)
and for other two arbitrary Hamiltonians $\widehat{\mathcal{H}}_{1}$
and $\widehat{\mathcal{H}}_{2}$ with parameters ($b=0.5,$ $a=1$)
and ($b=0.5,$ $a=1.8$) respectively. By doing a linear fitting (solid
line), we obtain \begin{equation}
A=(0.49\pm0.08)\text{ and\textrm{ }}B=(1.00\pm0.06)\label{eq:A&B}\end{equation}
%


It is important to note that the role of the interaction nature on
the decoherence time is manifested through the constants $A$ and
$B$. The origin of this factors will be clarified in the next section
by using the Keldysh form of the quantum field many-body theory. There,
a microscopic model is applied to a similar spin system. We will obtain
analytical expressions that agree well with the numerical values obtained
here.

\section{\textbf{Two-spin channel coupled to a spin bath:} Analytical solution\label{Analytical_Solution}}

\subsection{System}

\begin{figure}[tbh]
\begin{centering}
\includegraphics[scale=0.45]{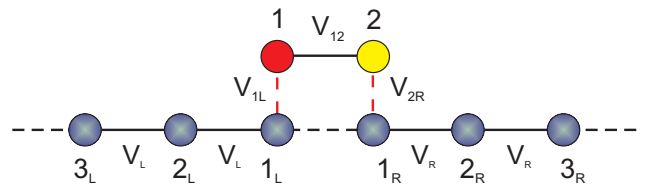}
\par\end{centering}

\caption{(Color online): Schematic representation of the two spin channel coupled
to independent spin environments.}

\centering{}\label{f--figure6}
\end{figure}

In this section, we consider the simplest quantum spin channel: a
two spin system that could act as a SWAP gate. Each of these spins
is coupled with an independent spin environment as is sketched in
Fig. \ref{f--figure6}. Since the system describes a Rabi oscillation
that could be damped by the interaction of the spin bath, it contains
all the essential ingredients of the spin channel case, with the further
advantage that it could be solved analytically and to asses in detail
how the environment disturbs it. Therefore, it allows the exact characterization
of the decoherence rate in order to compare with the numerical results
of the previous section.

The Hamiltonian of the two spin system is given by $\widehat{\mathcal{H}}_{\mathrm{I}}$
of Eq. (\ref{eq--Chain_Hamiltonian}) with $M=2$. The system-environment
Hamiltonian is equivalent to Eq. (\ref{eq--Tranverse H}) with $M=2$.
Finally, the environment is represented by two independent and semi-infinite
linear chains whose Hamiltonians are given by $\widehat{\mathcal{H}}_{\mathrm{II}}$
of Eq. (\ref{eq--Chain_Hamiltonian}) with $M\rightarrow\infty.$
By neglecting the interaction between the semi-infinite portions we
ensure that no correlation could appear. We call them $\widehat{\mathcal{H}}_{\mathrm{EL}}$
and $\widehat{\mathcal{H}}_{\mathrm{ER}},$ see figure \ref{f--figure6}.
As the theoretical framework we adopt the Keldysh formalism \cite{z--Danielewicz}
in a form that was developed for electronic excitations \cite{z--GLBEI,z--GLBEII}
and then extensively developed to solve the spin dynamics in presence
of system-environment interactions \cite{z-SPC,z--NuestroCPL05,z--NuestroSSC07,z--NuestroPRA07}.
More recently, this formalism also proved to be useful to address
thermal transport \cite{Arrachea-Lozano-Aligia}.

In this work, we briefly discuss the main points of the formalism
required to arrive to the solution. We start by establishing the relation
between spin and fermion operators at a given site $n$ by applying
the Jordan-Wigner transformation\ (JWT) \cite{Lieb61} \begin{equation}
\widehat{S}_{n}^{+}=\widehat{c}_{n}^{+}\exp\{\mathrm{i}\pi\sum_{m=1}^{n-1}\widehat{c}_{m}^{+}\widehat{c}_{m}^{{}}\}.\end{equation}
 Here, $\widehat{c}_{n}^{+}$ and $\widehat{c}_{n}^{{}}$ stand for
the creation and destruction fermionic operators, and $\widehat{S}_{n}^{\pm}$
are the rising and lowering spin operator, $\widehat{S}_{n}^{\pm}=\widehat{S}_{n}^{x}\pm\mathrm{i}\widehat{S}_{n}^{y}$.
Within the fermionic description, the spin-up and spin-down states
correspond to an occupied and not-occupied fermionic state respectively.
%

After applying the JWT to the total two spin Hamiltonian\begin{equation}
\widehat{\mathcal{H}}^{2\text{-}\mathrm{spin}}=\widehat{\mathcal{H}}_{\mathrm{I}}+\widehat{\mathcal{H}}_{\mathrm{T}}+\widehat{\mathcal{H}}_{\mathrm{EL}}+\widehat{\mathcal{H}}_{\mathrm{ER}},\end{equation}
it is possible to rewrite the different contributions of $\widehat{\mathcal{H}}^{2\text{-\textrm{spin}}}$.
$\widehat{\mathcal{H}}_{\mathrm{I}}$ is given by\begin{equation}
\widehat{\mathcal{H}}_{\mathrm{I}}=V_{12}\left(\hat{c}_{1}^{+}\hat{c}_{2}^{{}}+\hat{c}_{2}^{+}\hat{c}_{1}^{{}}\right),\end{equation}
 where the hopping amplitude between states $1$ and $2$ is \[
V_{12}\equiv J/2\]
 and the transition between them occurs at the natural Rabi frequency
\begin{equation}
\omega_{0}=2V_{12}/\hbar.\label{natRABI}\end{equation}
 The environment is represented by \begin{multline}
\widehat{\mathcal{H}}_{\mathrm{EL}}+\widehat{\mathcal{H}}_{\mathrm{ER}}=\sum_{i=1}^{\infty}V_{\mathrm{L}}^{{}}\left(\hat{c}_{\mathrm{L,}i}^{+}\hat{c}_{\mathrm{L},i+1}^{{}}+\hat{c}_{\mathrm{L},i+1}^{+}\hat{c}_{\mathrm{L},i}^{{}}\right)\\
+\sum_{i=1}^{\infty}V_{\mathrm{R}}^{{}}\left(\hat{c}_{\mathrm{R},i}^{+}\hat{c}_{\mathrm{R},i+1}^{{}}+\hat{c}_{\mathrm{R},i+1}^{+}\hat{c}_{\mathrm{R},i}^{{}}\right),\end{multline}
 where $\hat{c}_{\mathrm{L},i}^{+}(\hat{c}_{\mathrm{L},i}^{{}})$
is the creation (destruction) fermionic operator for the environment
that interacts with the system state $1$ and similarly for $\hat{c}_{\mathrm{R},i}^{+}(\hat{c}_{\mathrm{R},i}^{{}})$
which belongs to the environment that interacts with the system state
$2$. $V_{\mathrm{L}}$ and $V_{\mathrm{R}}$ stand for the hopping
term between neighboring sites within each of the environments. They
correspond to the XY (flip-flop) interaction along the $x$ direction.
\begin{equation}
V_{\mathrm{L}}=V_{\mathrm{R}}\equiv J_{\mathrm{x}}/2~.\label{Jx}\end{equation}
 Finally, the transversal or system-environment interaction takes
the form \begin{multline}
\widehat{\mathcal{H}}_{\mathrm{T}}=V_{1\mathrm{L}}^{\mathrm{XY}}~(\hat{c}_{\mathrm{L},1}^{+}\hat{c}_{1}^{{}}+\hat{c}_{1}^{+}\hat{c}_{\mathrm{L},1}^{{}})\\
+V_{1\mathrm{L}}^{\mathrm{ZZ}}~\left[\hat{c}_{\mathrm{L},1}^{+}\hat{c}_{\mathrm{L},1}^{{}}\hat{c}_{1}^{+}\hat{c}_{1}^{{}}-\frac{1}{2}\hat{c}_{\mathrm{L},1}^{+}\hat{c}_{\mathrm{L},1}^{{}}-\frac{1}{2}\hat{c}_{1}^{+}\hat{c}_{1}^{{}}+\frac{1}{4}\right]\\
+V_{2\mathrm{R}}^{\mathrm{XY}}~(\hat{c}_{\mathrm{R},1}^{+}\hat{c}_{2}^{{}}+\hat{c}_{2}^{+}\hat{c}_{\mathrm{R},1}^{{}})\\
+V_{2\mathrm{R}}^{\mathrm{ZZ}}~\left[\hat{c}_{\mathrm{R},1}^{+}\hat{c}_{\mathrm{R},1}^{{}}\hat{c}_{2}^{+}\hat{c}_{2}^{{}}-\frac{1}{2}\hat{c}_{\mathrm{R},1}^{+}\hat{c}_{\mathrm{R},1}^{{}}-\frac{1}{2}\hat{c}_{2}^{+}\hat{c}_{2}^{{}}+\frac{1}{4}\right],\label{eq--SE_inter_fermion}\end{multline}
 where \begin{align*}
V_{1\mathrm{L}}^{\mathrm{XY}} & =bJ_{1\mathrm{L}}/2\\
V_{2\mathrm{R}}^{\mathrm{XY}} & =bJ_{2\mathrm{R}}/2\end{align*}
 are the hopping amplitudes, due to the $XY$ term of the interaction,
between the site $1$ and $2$ in the quantum channel with the left
\ and right environments respectively. The standard direct integral
of the Coulomb interaction of an electron (fermion) in state $1$
(2) with an electron in the first site of the left (right) reservoir
correspond to an Ising interaction between spins \begin{align*}
V_{1\mathrm{L}}^{\mathrm{ZZ}} & =aJ_{1\mathrm{L}}\equiv aJ_{\mathrm{y}}\\
V_{2\mathrm{R}}^{\mathrm{ZZ}} & =aJ_{2\mathrm{R}}\equiv aJ_{\mathrm{y}}\end{align*}
 Analogously, the XY (flip-flop) component of the inter-chain interaction
is associated to the hopping amplitudes\begin{equation}
V_{1\mathrm{L}}^{\mathrm{XY}}=V_{2\mathrm{R}}^{\mathrm{XY}}\equiv bJ_{\mathrm{y}}/2.\end{equation}
 Note that the Ising term of the system-environment interaction in
the spin problem is not completely analogous to the Coulomb interaction
within the fermionic description, since after the JWT the exchange
term of the Coulomb interaction is not present. The last three terms
in each of the brackets of Eq. (\ref{eq--SE_inter_fermion}) do not
involve an interaction between the system and the environment. Thus
they just modify the potential energies of sites $1$, $2$ and site
$1$ of both environments. Since the potential energy can be controlled
externally to maximize the polarization transfer in a NMR experiment
\cite{z--HartmannHahn,z--Nuestro JCP2006}, these terms will be neglected
for the present calculations.

\subsection{Spin dynamics within the Keldysh formalism}

Even at room temperature $k_{\mathrm{B}}T$ is much\ higher than
any energy scale involve in an NMR experiment. The local polarization
given by Eq. (\ref{eq--PCFunc}) can be expressed within the Keldysh
formalism as \cite{z-SPC}\begin{equation}
P_{m,n}(t)=\tfrac{2\hbar}{\mathrm{i}}G_{m,m}^{<\,}(t,t)-1.\label{eq--PCFunc_Keldysh}\end{equation}
 Here, $G_{m,m}^{<\,}(t,t)$ is a particular case of the general particle
density function \cite{z--Keldysh} \begin{equation}
G_{m,n}^{<}(t_{2},t_{1})=\tfrac{\mathrm{i}}{\hbar}\left\langle \Psi_{\mathrm{n.e.}}\right\vert \widehat{c}_{m}^{+}(t_{1})\widehat{c}_{n}^{{}}(t_{2})\left\vert \Psi_{\mathrm{n.e.}}\right\rangle .\end{equation}
 The initial polarized state is described by the non-equilibrium state
$\left\vert \Psi_{\mathrm{n.e.}}\right\rangle =\widehat{c}_{n}^{+}\left\vert \Psi_{\mathrm{eq.}}\right\rangle $
formed by creating an excitation at $t=0$ on the $n$-th site. Therefore,
the non-equilibrium density $G_{m,m}^{<\,}(t,t)$ of Eq. (\ref{eq--PCFunc_Keldysh})
depends implicitly on the index $n$ that indicates the site of the
initial excitation. The expression for this initial condition can
be expressed as\begin{equation}
G_{k,l}^{<}(0,0)=\tfrac{\mathrm{i}}{2\hbar}\left(\delta_{k,l}+2\Delta P~\delta_{k,n}\delta_{n,l}\right),\label{eq--Initial density_}\end{equation}
 where the first term describes the equilibrium density which is identical
for all sites and does not contribute to the dynamics. The second
term represents the non-equilibrium contribution where only for the
$n$-th site is different from zero. Notice that this initial state
represents a non-correlated initial state, where the only elements
different from zero are those satisfying $k=l$. The factor $\Delta P$
accounts for the excess of excitation at site $n$ and is the responsible
of the observed dynamics. Its value ranges from zero (lack of excitation)
to a maximum of $1/2.$

The initial density function, Eq. (\ref{eq--Initial density_}), evolves
under the Schrödinger equation that could be expressed in the Danielewicz
form \cite{z--Danielewicz} given by the expression \begin{multline}
\mathbf{G}^{<}\left(t_{2},t_{1}\right)=\hbar^{2}\mathbf{G}^{\mathrm{R}}\left(t_{2},0\right)\mathbf{G}^{<}\left(0,0\right)\mathbf{G}^{\mathrm{A}}\left(0,t_{1}\right)+\\
\int_{0}^{t_{2}}\int_{0}^{t_{1}}\mathrm{d}t_{k}\mathrm{d}t_{l}\mathbf{G}^{\mathrm{R}}\left(t_{2},t_{k}\right)\mathbf{\Sigma}^{<}\left(t_{k},t_{l}\right)\mathbf{G}^{\mathrm{A}}\left(t_{l},t_{1}\right).\label{Danielewicz_evol}\end{multline}
 Here, $\mathbf{G}^{<}\left(t_{2},t_{1}\right)$ is a $2\times2$
density matrix whose elements $G_{m,n}^{<}(t_{2},t_{1})$ are restricted
to $m,n\in\left\{ 1,2\right\} $. Similarly, $\mathbf{G}^{\mathrm{R}}\left(t_{2},t_{1}\right)$
represents an effective evolution operator in this reduced space whose
elements, the retarded Green's function \begin{align}
G_{m,n}^{\mathrm{R}}\left(t_{2},t_{1}\right) & =\left[G_{n,m}^{\mathrm{A}}\left(t_{1},t_{2}\right)\right]^{\dagger}\\
 & =\theta\left(t_{2},t_{1}\right)\ [G_{m,n}^{>}\left(t_{2},t_{1}\right)-G_{m,n}^{<}\left(t_{2},t_{1}\right)],\end{align}
 describe the probability of finding an excitation at site $m$ after
it was placed at site $n$ and evolved under the total Hamiltonian
for a time $t_{2}-t_{1}$. The injection self-energy, $\mathbf{\Sigma}^{<}$,
takes into account the effects of the environment. The first term
of Eq. (\ref{Danielewicz_evol}) stands for the {}``coherent\textquotedblright{}
evolution since it preserves the memory of the initial excitation,
while the second term contains {}``incoherent reinjections\textquotedblright{}
described by the injection self-energy that compensates any leak from
the coherent evolution \cite{z--GLBEII}.

In absence of $\widehat{\mathcal{H}}_{\mathrm{T}}$, the retarded
Green's function for the system is easily evaluated in its energy
representation given by the following expression\begin{equation}
\mathbf{G}^{0\mathrm{R}}\left(\varepsilon\right)=\int\mathbf{G}^{0\mathrm{R}}\left(t\right)\exp[\mathrm{i}\varepsilon t/\hbar]\mathrm{d}t=[\varepsilon\mathbf{I-H}_{\mathrm{I}}]^{-1}.\end{equation}
 Conversely, with presence of $\widehat{\mathcal{H}}_{\mathrm{T}}$,
the interacting Green's function defines the reduced effective Hamiltonian
\begin{equation}
\mathbf{H}_{\mathrm{eff.}}(\varepsilon)\equiv\varepsilon\mathbf{I}-\left[\mathbf{G}^{\mathrm{R}}\left(\varepsilon\right)\right]^{-1}=\mathbf{H}_{\mathrm{I}}+\mathbf{\Sigma}^{\mathrm{R}}(\varepsilon)\end{equation}
 and the self-energies $\mathbf{\Sigma}^{\mathrm{R}}(\varepsilon)$
\cite{z--DAmato}, where the exact perturbed dynamics is contained
in the nonlinear dependence of the self-energies $\Sigma^{\mathrm{R}}$
on $\varepsilon$. For infinite reservoirs, $\operatorname{Re}\Sigma^{\mathrm{R}}\left(\varepsilon_{\nu}\right)=\operatorname{Re}\left(\varepsilon_{\nu}-\varepsilon_{\nu}^{o}\right)$
represents the {}``shift\textquotedblright{}\ of the system's eigen-energies
$\varepsilon_{\nu}^{o}$ with $\varepsilon_{\nu}$ the eigen-energies
of the interacting problem. The imaginary part of $\Sigma^{\mathrm{R}}$,\begin{equation}
-2\operatorname{Im}\Sigma^{\mathrm{R}}\left(\varepsilon_{\nu}\right)/\hbar=1/\tau_{\mathrm{T}}^{{}}=2\Gamma_{\mathrm{T}}/\hbar,\label{eq--tauSE_definition}\end{equation}
 accounts for their\ {}``decay rate\textquotedblright{} into collective
system-environment eigenstates in agreement with the Self-Consistent
Fermi Golden Rule \cite{z--elenaCPL05}, i.e. the evolution with $\mathbf{H}_{\mathrm{eff.}}$
is non-unitary.

We use a perturbative expansion on $\widehat{\mathcal{H}}_{\mathrm{T}}$
to build up expressions for the particle (hole) self-energies $\mathbf{\Sigma}^{<(>)}(t_{k},t_{l})$
as well as for the retarded (advanced) self-energies $\mathbf{\Sigma}^{\mathrm{R(A)}}(t_{k},t_{l})$.
Under the wide band assumption (or fast fluctuation approximation),
where the dynamics of excitations within the environments are faster
than the relevant time scales of the system $V_{\mathrm{L}},V_{\mathrm{R}}\gg V_{12}$,
we obtain for the decay rates of the $XY$ and Ising system-environment
interaction \cite{z--NuestroCPL05,z--NuestroSSC07,z--NuestroPRA07}
the following expressions \begin{align}
\frac{2}{\hbar}\Gamma_{1\mathrm{T}}^{\mathrm{XY}} & =\frac{2\pi}{\hbar}\left\vert V_{1\mathrm{L}}^{\mathrm{XY}}\right\vert ^{2}\frac{1}{\pi V_{\mathrm{L}}},\label{eq--decayrateXY1}\\
\frac{2}{\hbar}\Gamma_{2\mathrm{T}}^{\mathrm{XY}} & =\frac{2\pi}{\hbar}\left\vert V_{2\mathrm{R}}^{\mathrm{XY}}\right\vert ^{2}\frac{1}{\pi V_{\mathrm{R}}},\label{eq--decayrateXY2}\end{align}
 and\begin{align}
\frac{2}{\hbar}\Gamma_{1\mathrm{T}}^{\mathrm{ZZ}} & =\frac{2\pi}{\hbar}\left\vert V_{1\mathrm{L}}^{\mathrm{ZZ}}\right\vert ^{2}\frac{8(\frac{1}{4}-\Delta P_{1}^{2})}{3\pi^{2}V_{\mathrm{L}}},\label{eq--decayrateI1}\\
\frac{2}{\hbar}\Gamma_{2\mathrm{T}}^{\mathrm{ZZ}} & =\frac{2\pi}{\hbar}\left\vert V_{2\mathrm{R}}^{\mathrm{ZZ}}\right\vert ^{2}\frac{8(\frac{1}{4}-\Delta P_{2}^{2})}{3\pi^{2}V_{\mathrm{R}}}.\label{eq--decayrateI2}\end{align}
 Here, 
it is considered that the left and right environments are in equilibrium.
Thus, the occupation for any of their sites is $\mathrm{f}_{\mathrm{L,R}}=\left(1/2+\Delta P_{1,2}\right),$
with $\Delta P_{1,2}$ representing the occupation excess on the left
and right environment respectively. Due to the wide band approximation
the decay rates are time and energy independent, and in consequence
$\mathbf{H}_{\mathrm{eff.}}$ does not depend on $\varepsilon$ \begin{align*}
\mathbf{H}_{\mathrm{eff.}} & =\left(\begin{array}{cc}
0 & -V_{12}\\
-V_{12} & 0\end{array}\right)+\left(\begin{array}{cc}
-\mathrm{i}\Gamma_{1\mathrm{T}} & 0\\
0 & -\mathrm{i}\Gamma_{2\mathrm{T}}\end{array}\right)\\
 & =\left(\begin{array}{cc}
-\mathrm{i}\Gamma_{1\mathrm{T}} & -V_{12}\\
-V_{12} & -\mathrm{i}\Gamma_{2\mathrm{T}}\end{array}\right),\end{align*}
 where $\Gamma_{1\mathrm{T}}=\Gamma_{1\mathrm{T}}^{\mathrm{XY}}+\Gamma_{\mathrm{1T}}^{\mathrm{ZZ}}$
and $\Gamma_{2\mathrm{T}}=\Gamma_{2\mathrm{T}}^{\mathrm{XY}}+\Gamma_{2\mathrm{T}}^{\mathrm{ZZ}}$.
In order to obtain a comparison with the numerical results of the
previous section, we assume that the excess of occupation in the left
(right) environment $\Delta P_{1}$ $\left(\Delta P_{2}\right)$ is
very small ($\Delta P_{1},\Delta P_{2}\ll\frac{1}{2}$), condition
that is well satisfied in the high temperature regime, and that the
hopping amplitudes satisfy $V_{1\mathrm{L}}^{\mathrm{ZZ}}=V_{2\mathrm{R}}^{\mathrm{ZZ}}$,
$V_{1\mathrm{L}}^{\mathrm{XY}}=V_{2\mathrm{R}}^{\mathrm{XY}}$ and
$V_{\mathrm{L}}=V_{\mathrm{R}}$. They ensure that the decay rates
to the left and right environments are identical, i.e. $\Gamma_{1\mathrm{T}}=\Gamma_{2\mathrm{T}}=\Gamma_{\mathrm{T}}$.
Under these conditions, the propagator has a simple dependence on
$t$ given by $\mathbf{G}^{\mathrm{R}}\left(t\right)=\mathbf{G}^{0\mathrm{R}}\left(t\right)e^{-\Gamma_{\mathrm{T}}t/\hbar}$,
where $G_{11}^{0\mathrm{R}}(t)=G_{22}^{0\mathrm{R}}(t)=\frac{\mathrm{i}}{\hbar}\cos\left(\frac{\omega_{0}}{2}t\right)$
and $G_{12}^{0\mathrm{R}}(t)=G_{21}^{0\mathrm{R}}(t)=\frac{\mathrm{i}}{\hbar}\sin\left(\frac{\omega_{0}}{2}t\right)$.
Then Eq. (\ref{Danielewicz_evol}) becomes \begin{multline}
\mathbf{G}^{<}\left(t,t\right)=\hbar^{2}\mathbf{G}^{0\mathrm{R}}\left(t\right)\mathbf{G}^{<}\left(0,0\right)\mathbf{G}^{0\mathrm{A}}\left(-t\right)e^{-t/\tau_{\mathrm{T}}}+\\
\int_{0}^{t}\mathrm{d}t_{i}\mathbf{G}^{0\mathrm{R}}\left(t-t_{i}\right)\mathbf{\Sigma}^{<}\left(t_{i}\right)\mathbf{G}^{0\mathrm{A}}\left(t_{i}-t\right)e^{-\left(t-t_{i}\right)/\tau_{\mathrm{T}}},\label{danielewicz2}\end{multline}
 which is a generalized Landauer-Büttiker equation \cite{z--GLBEI,z--GLBEII}.
This equation is complemented with the injection self-energy $\mathbf{\Sigma}_{{}}^{<},$
which takes into account for particles that return to the system after
an interaction with the environment, and is expressed as \cite{z--NuestroCPL05,z--NuestroSSC07,z--NuestroPRA07}
\begin{align}
\Sigma_{ij}^{<} & =\left[2\Gamma_{\mathrm{T}}^{\mathrm{ZZ}}\hbar G_{ii}^{<}+\hbar2\Gamma_{\mathrm{T}}^{\mathrm{XY}}\frac{\mathrm{i}}{\hbar}\left(\frac{1}{2}+\Delta P_{i}\right)\right]~\delta_{ij}\label{eq--Sigma<}\\
 & \cong\left[2\Gamma_{\mathrm{T}}^{\mathrm{ZZ}}\hbar G_{ii}^{<}+\hbar2\Gamma_{\mathrm{T}}^{\mathrm{XY}}\frac{\mathrm{i}}{\hbar}\frac{1}{2}\right]~\delta_{ij},\end{align}
 where in the last expression we used the assumption $\Delta P_{1},\Delta P_{2}\ll1/2.$

\subsection{Characterization of the decoherent processes}

We solve Eq. (\ref{danielewicz2}) together with the injection self-energy
of Eq. (\ref{eq--Sigma<}) with an initial condition given by an excitation
of $\Delta P$ on site $1$, i.e. $\frac{\hbar}{\mathrm{i}}G_{ij}^{<}\left(0,0\right)=\Delta P\delta_{i1}\delta_{1j}.$
For this purpose we follow the strategy used in Refs. \cite{z--NuestroSSC07,z--NuestroPRA07}.
Replacing Eq. (\ref{eq--Sigma<}) into Eq. (\ref{danielewicz2}) and\ identifying
the interaction rate of Eq. (\ref{eq--tauSE_definition}), we get
two coupled equations for $G_{11}^{<}$ and $G_{22}^{<}$ given by\begin{gather}
\tfrac{\hbar}{\mathrm{i}}G_{\genfrac{}{}{0pt}{}{11}{(22)}}^{<}\left(t,t\right)=\hbar^{2}\left\vert G_{\genfrac{}{}{0pt}{}{11}{(21)}}^{0\mathrm{R}}\left(t\right)\right\vert ^{2}\Delta Pe^{-t/\tau_{\mathrm{T}}}\nonumber \\
+\int\left\vert G_{\genfrac{}{}{0pt}{}{11}{(21)}}^{0\mathrm{R}}\left(t-t_{i}\right)\right\vert ^{2}e^{-(t-t_{i})/\tau_{\mathrm{T}}}\left[\tfrac{\hbar}{\mathrm{i}}\Sigma_{11}^{<}\left(t_{i}\right)\right]\mathrm{d}t_{i}\nonumber \\
+\int\left\vert G_{\genfrac{}{}{0pt}{}{12}{(22)}}^{0\mathrm{R}}\left(t-t_{i}\right)\right\vert ^{2}e^{-(t-t_{i})/\tau_{\mathrm{T}}}\left[\tfrac{\hbar}{\mathrm{i}}\Sigma_{22}^{<}\left(t_{i}\right)\right]\mathrm{d}t_{i}.\label{eq--GLBE-_1and 2}\end{gather}
 The first term is the probability that a particle, initially on site
$1$, is found at time $t$ on site $1$ (or $2$) having survived
the interaction with the environment. The second and third terms describe
particles whose last interaction with the environment, at time $t_{i}$,
occurred at site $1$ and $2$ respectively. Noticeably, in the first
term of Eq. (\ref{eq--GLBE-_1and 2}) the environment, though giving
the exponential decay, does not affect the frequency of the two spin
system given in $G_{\genfrac{}{}{0pt}{}{11}{(21)}}^{0\mathrm{R}}\left(t\right)$.
Modification of $\omega$ requires the dynamical feedback contained
in the next terms. The solution of Eq. (\ref{eq--GLBE-_1and 2}) involves
a Laplace transform. This solution, for the general case where the
system-environment interaction involves both $XY$ and Ising terms,
gives for the local polarization of Eq. (\ref{eq--PCFunc_Keldysh})
the following expression \begin{multline}
P_{11}(t)=2\Delta P\left[\frac{1}{2}e^{-2\Gamma_{\mathrm{T}}^{\mathrm{XY}}t/\hbar}\right.\\
\left.+\frac{1}{2\cos\left(\phi\right)}\cos\left[\left(\omega+\mathrm{i}\eta\right)t+\phi\right]e^{-\left(2\Gamma_{\mathrm{T}}^{\mathrm{XY}}+\Gamma_{\mathrm{T}}^{\mathrm{ZZ}}\right)t/\hbar}\right],\label{G11}\end{multline}
 where \begin{align}
\omega & =\left\{ \begin{array}{cc}
\omega_{0}\sqrt{1-\left(\frac{\Gamma_{\mathrm{T}}^{\mathrm{ZZ}}}{\hbar\omega_{0}}\right)^{2}} & ~~~~\hbar\omega_{0}>\Gamma_{\mathrm{T}}^{\mathrm{ZZ}}\\
0 & ~~~~\hbar\omega_{0}\leq\Gamma_{\mathrm{T}}^{\mathrm{ZZ}}\end{array}\right.,\label{frequency}\\
\eta & =\left\{ \begin{array}{cc}
0 & ~~~~\hbar\omega_{0}>\Gamma_{\mathrm{T}}^{\mathrm{ZZ}}\\
\omega_{0}\sqrt{\left(\frac{\Gamma_{\mathrm{T}}^{\mathrm{ZZ}}}{\hbar\omega_{0}}\right)^{2}-1} & ~~~~\hbar\omega_{0}\leq\Gamma_{\mathrm{T}}^{\mathrm{ZZ}}\end{array}\right.,\label{relaxation_w}\end{align}
 and\begin{equation}
\tan\left(\phi\right)=-\frac{\Gamma_{\mathrm{T}}^{\mathrm{ZZ}}}{\hbar\omega}.\end{equation}
 From these expressions it is possible to determine the observable
frequency $\omega$ and the decoherence time as the slowest of the
two competing interaction rates: $\hbar/\left[\left(2\Gamma_{\mathrm{T}}^{\mathrm{XY}}+\Gamma_{\mathrm{T}}^{\mathrm{ZZ}}\right)\pm\eta\right]$.

It is interesting to remark that the effect of the lateral chains
on the two spin system coupled through an Ising system-environment
interaction can produce observables with non-linear dependences on
$\widehat{\mathcal{H}}_{\mathrm{T}}$. We find a \emph{non-analyticity}
in these functions enabled by the infinite degrees of freedom of the
environment \cite{z--sachdev} (i.e. the thermodynamic limit). Here,
they are incorporated through the respective imaginary part of the
self-energy, $\hbar/\Gamma_{\mathrm{T}}^{\mathrm{ZZ}},$ i.e. the
FGR. \ Hence, the non-analyticity of $\omega$ and $\tau_{\phi}$
on the control parameter $\hbar\omega_{0}/\Gamma_{\mathrm{T}}^{\mathrm{ZZ}}$
at the critical value $\hbar\omega_{0}/\Gamma_{\mathrm{T}}^{\mathrm{ZZ}}=1$,
indicates a switch between two dynamical regimes. In previous works,
we identified this behavior as a Quantum Dynamical Phase Transition
\cite{z--Nuestro JCP2006,z--NuestroSSC07}. This can be interpreted
as a disruption of the environment into the dynamical nature of the
system, a form of the Quantum Zeno Effect (QZE), which states that
quantum dynamics is slowed down by a frequent measurement process
\cite{QZE}. Very recently, this dynamical phase transition was evidenced
as a particular scale invariance in the fluctuations in the number
of photons emitted by a driven two level system \cite{Garraham-Lesanovsky}.
However, here we choose to work in the regime where decoherence rate
is still weak to produce such transitions. The evaluation of the conditions
to observe such dynamical phase transition or a cross-over in the
dynamics dimensionality \cite{Pastawski-Usaj} is an open problem
that deserves further studies.

For a pure Ising system-environment interaction, $\Gamma_{\mathrm{T}}=\Gamma_{\mathrm{T}}^{\mathrm{ZZ}}$,
i.e., $\Gamma_{\mathrm{T}}^{\mathrm{XY}}=0,$ we recover the expression
found in Ref. \cite{z--NuestroSSC07}. For a pure $XY$ system-environment
interaction $\Gamma_{\mathrm{T}}=\Gamma_{\mathrm{T}}^{\mathrm{XY}}$,
i.e., $\Gamma_{\mathrm{T}}^{\mathrm{ZZ}}=0,$ it is noticeably that
the observable frequency and decoherence time depend linearly on $\widehat{\mathcal{H}}_{\mathrm{T}}$.
In particular, the oscillation frequency coincides with that of the
isolated system, i.e. $\omega=\omega_{0}$. This is due to the symmetry
of the decay rates to both environments, $\Gamma_{1\mathrm{T}}=\Gamma_{2\mathrm{T}}$.
In absence of this symmetry this effect is not observed as can be
seen for a spin system interacting with only one environment \cite{z--NuestroCPL05,z--Nuestro JCP2006,z--NuestroPRA07}.

It is remarkably that in the limit $\hbar\omega_{0}\gg\Gamma_{\mathrm{T}}^{\mathrm{ZZ}}$,
the solution (\ref{G11}) tends to the solution of a two spin system
only coupled to the environment through one spin of the system \cite{PhysB}.
Thus, within this regime it is impossible to identify whether the
two spin system is coupled to two wide band environments or to a single
one.

From Eq. (\ref{G11}) it is straightforward to obtain the decoherence
time within the regime $J_{\mathrm{y}}\ll J_{\mathrm{x}}$ where $1/\tau_{\phi}=\left(\Gamma_{\mathrm{T}}^{\mathrm{ZZ}}+2\Gamma_{\mathrm{T}}^{\mathrm{XY}}\right)/\hbar$.
In order to compare with the numerical results of previous section
we identify the intra-chain and inter-chain hopping as\begin{equation}
V_{\mathrm{L}}=V_{\mathrm{R}}\equiv J_{\mathrm{x}}/2~\ \label{Vintra}\end{equation}
 \begin{equation}
\text{and \ \ }V_{1\mathrm{L}}^{\mathrm{XY}}=V_{2\mathrm{R}}^{\mathrm{XY}}\equiv bJ_{\mathrm{y}}/2,\label{Vxy}\end{equation}
 while the through space inter-chain Coulomb coupling results: \begin{equation}
V_{1\mathrm{L}}^{\mathrm{ZZ}}=V_{2\mathrm{R}}^{\mathrm{ZZ}}\equiv aJ_{\mathrm{y}}.\label{Vzz}\end{equation}
 Thus, taking in mind the high temperature regime $\left(\Delta P_{1},\Delta P_{2}\ll\frac{1}{2}\right),$
we obtain for the decay rates $\Gamma_{\mathrm{T}}^{\mathrm{ZZ}}$
and $\Gamma_{\mathrm{T}}^{\mathrm{XY}}$ the following values\begin{equation}
\frac{2}{\hbar}\Gamma_{\mathrm{T}}^{\mathrm{XY}}=\frac{2\pi}{\hbar}\frac{1}{2}\left\vert bJ_{\mathrm{y}}\right\vert ^{2}\frac{1}{\pi J_{\mathrm{x}}}\label{XY_rate_comparison}\end{equation}
 and\begin{equation}
\frac{2}{\hbar}\Gamma_{\mathrm{T}}^{\mathrm{ZZ}}=\frac{2\pi}{\hbar}\left\vert aJ_{\mathrm{y}}\right\vert ^{2}\frac{4}{3\pi^{2}J_{\mathrm{x}}}.\label{Ising_rate_comparison}\end{equation}
 It is interesting to note that these two decay rates are not equal
for an isotropic system-environment interaction $\left(a/b=1\right)$
\cite{z--NuestroPRA07} where its ratio is given by\begin{equation}
\frac{\Gamma_{\mathrm{T}}^{\mathrm{ZZ}}}{\Gamma_{\mathrm{T}}^{\mathrm{XY}}}=\frac{8}{3\pi}\approx0.849.\label{ratio_XY_ISING}\end{equation}
 Moreover, increasing the occupation of the environments $\Delta P_{1}$
and $\Delta P_{2}$ the ratio is reduced. This result contrasts with
the usual assumption of taking them equal \cite{z--QmasterE1,z--QmasterE2,MKBE74,JCP03,z--Nuestro JCP2006}.
Thus, here we show that always an $XY$ system-environment interaction
is more effective than the Ising one to destroy coherence. Observing
the decoherence rate $1/\tau_{\phi}=\left(\Gamma_{\mathrm{T}}^{\mathrm{ZZ}}+2\Gamma_{\mathrm{T}}^{\mathrm{XY}}\right)/\hbar,$
an extra factor of $1/2$ reduce the Ising decoherence rate. In addition,
one can decrease the Ising decoherence process by increasing the occupation
within the reservoir.

By using Eqs. (\ref{XY_rate_comparison}) and (\ref{Ising_rate_comparison})
we obtain for the decoherence rate\begin{align}
1/\tau_{\phi} & =\frac{1}{\tau_{\phi}^{\mathrm{ZZ}}}+\frac{1}{\tau_{\phi}^{\mathrm{XY}}}\label{Jy2/Jx}\\
 & =\left(\Gamma_{\mathrm{T}}^{\mathrm{ZZ}}+2\Gamma_{\mathrm{T}}^{\mathrm{XY}}\right)/\hbar=\frac{4}{3\pi}\frac{a^{2}\left\vert J_{\mathrm{y}}\right\vert ^{2}}{\hbar J_{\mathrm{x}}}+\frac{b^{2}\left\vert J_{\mathrm{y}}\right\vert ^{2}}{\hbar J_{\mathrm{x}}}\\
 & \approx0.424\frac{a^{2}\left\vert J_{\mathrm{y}}\right\vert ^{2}}{\hbar J_{\mathrm{x}}}+\frac{b^{2}\left\vert J_{\mathrm{y}}\right\vert ^{2}}{\hbar J_{\mathrm{x}}}\end{align}
 These analytical results are in notable agreement with those obtained
from the numerical solutions of the dynamics of spin systems computed
in the preceding section, Eqs. (\ref{eq--taum1Iandtaum1XY}) and (\ref{eq:A&B}).
In particular it confirms a notable effect repeatedly observed experimentally:
a stronger interaction along the chain results in a weaker decoherence
\cite{Rufeil-Fiori-MQC}. Indeed, Eq. \ref{Jy2/Jx} states that a
fast in-chain dynamics makes the already slow inter-chain dynamics
even slower. This is a form of the QZE that is manifested experimentally
in the spin diffusion in low-dimensional crystals. Slightly different
crystals showed an unexpected dimensional crossover as a function
of a structural parameter \cite{Levstein-Spin-Diffusion} . This crossover
was described as a QZE where the internal degrees of freedom act as
measurement apparatus \cite{Pastawski-Usaj}. The concept that the
measurement is played by an interaction with another quantum object,
or simply another degree of freedom of the subsystem investigated,
was independently and fully formalized by recasting it in terms of
an adiabatic theorem by Pascazio and collaborators \cite{Pascazio-QZE}.
Conversely, a strong inter-chain interaction can even lead to a freeze
of the SWAP dynamics as\ described by Eq. \ref{frequency} and fully
characterized experimentally \cite{z--Nuestro JCP2006}.

Getting more into\ more precise details, the Keldysh description
of the spin dynamics enabled to obtain from first principles the origin
of the coefficients $A$ and $B$ obtained in previous section on
numerical grounds. While the precise scale depends on the success
of the model to yield a representative local density of directly connected
states, the relation between A and B depends on the nature of the
XY and Ising interaction and hence has a universal meaning 

\section{Conclusion}

We have studied numerically the quantum evolution of a localized initial
excitation within a spin-chain channel weakly coupled ($\left\vert J_{\mathrm{y}}/J_{\mathrm{x}}\right\vert \ll1$)
to a second lateral spin-chain, i.e. a spin ladder system. We characterized
the decoherence rate of the channel by measuring the attenuation of
mesoscopic echoes with respect to the isolated channel evolution for
different kinds of the inter-chain interactions. We showed that the
$XY$ system-environment interaction is more effectively to destroy
quantum coherences than the Ising system-environment interaction.
Notably,\ by increasing the environment occupation, $\Delta P_{1,2}$,
one can reduce even more the Ising decoherence rate.

The decoherence characterization was possible by resorting to the
analytical solution of a two spin channel, where each of the spins
are coupled to independent spin environments in a fast fluctuation
regime. That was done by using the Keldysh formalism. We showed that
a quantum dynamical phase transition only appear when an Ising system-environment
interaction is present. Thus, for an XY system-environment interaction
the bare two-spin oscillation frequency holds. The very good agreement
between the decoherence rates of the spin-chain channel and the two
spin system confirms that, within the weakly coupled regime, the finite
lateral chain behaves as an environment in a fast fluctuation regime.
Within this regime, in consequence the complex many-body evolution
of the spin ladder can be obtained from a one-body dynamics plus an
exponential decoherence process dictated by the Fermi Golden Rule.
Although simple in nature, this statement has important experimental
consequences. It helps to explain in terms of the QZE a notable experimental
observation in low dimensional spin dynamics: the stronger interaction
along the chain dominates over the weaker inter-chain inducing a stability
of one-dimensional dynamics against perturbations by spins outside
the chain \cite{Levstein-Spin-Diffusion,Rufeil-Fiori-MQC}. A natural
suggestion is that in a quantum channel one can use the attenuation
of mesoscopic echoes as a tool for characterizing decoherence in the
channel without using a receiver at the other end.
\begin{acknowledgments}
We acknowledge support from Fundaci\'{o}n Antorchas, CONICET, FoNCyT,
and SeCyT-UNC. G.A.A. and E.P.D. thank the Alexander von Humboldt
Foundation for a Research Scientist Fellowship. P.R.L. and H.M.P.
are members of the Research Career of CONICET. \end{acknowledgments}

\end{document}